# Using Your Beam Efficiently: Reducing Electron-dose in the STEM via Flyback Compensation

Tiarnan Mullarkey[1,2], Jonathan J. P. Peters[1], Clive Downing[3], Lewys Jones[1,3]

1. School of Physics, Trinity College Dublin, Dublin 2, Ireland.
2. Centre for Doctoral Training in the Advanced Characterisation of Materials, AMBER Centre, Dublin 2, Ireland.
3. Advanced Microscopy Laboratory, Centre for Research on Adaptive Nanostructures & Nanodevices (CRANN), Dublin 2, Ireland.

**Abstract**

In the scanning transmission electron microscope, fast-scanning and frame-averaging are two widely used approaches for reducing electron-beam damage and increasing image signal-noise ratio which require no additional specialised hardware. Unfortunately, for scans with short pixel dwell-times (less than 5 μs), line flyback time represents an increasingly wasteful overhead. Although beam exposure during flyback causes damage while yielding no useful information, scan-coil hysteresis means that eliminating it entirely leads to unacceptably distorted images. In this work, we reduce this flyback to an absolute minimum by calibrating and correcting for this hysteresis in postprocessing. Substantial improvements in dose-efficiency can be realised (up to 20 %), while crystallographic and spatial fidelity is maintained for displacement/strain measurement.

## Introduction

The scanning transmission electron microscope (STEM) is a powerful instrument for probing materials down to the atomic scale. However, reliably imaging fragile samples requires the operator to consider the risk of electron-beam damage introducing artefacts into the data. Many approaches have been introduced to reduce the electron-fluence, measured in $e^-/Å^2$, (often referred to as electron-dose in the electron microscopy community, which is the term used hereafter) or dose-rate ($e^-/Å^2\ s^{-1}$) imposed on the sample (Chen et al., 2020), or to increase a sample's durability with either fixation or cryogenic cooling (Glaeser, 1971; Elbaum, 2018). Recently, novel approaches such as compressed sensing (CS) have been proposed to reduce the fraction of pixels illuminated and so reduce the electron exposure (Anderson et al., 2013; Stevens et al., 2014, 2018). However, CS requires additional fast (and expensive) hardware to be introduced that may not be available to all microscopists (Kovarik et al., 2016; Béché et al., 2016). Moreover, where readout noise is not the limiting factor, such as with digital read-out approaches (Jones & Downing, 2018; Mullarkey et al., 2020) CS may not offer any appreciable benefit over conventional (Shannon) scanning (Sanders & Dwyer, 2018; Van den Broek et al., 2019).

Instead, approaches which rely only on reducing beam-current (Buban et al., 2010), reducing scanning density (more coarse scans) (Yankovich et al., 2015), or reducing dwell-time (Mittelberger et al., 2018) may be attractive as they are available to all STEM operators without the purchase of additional hardware. With these approaches, individual frames may then exhibit a relatively poor signal-noise ratio (SNR) but a final image is reconstructed through frame alignment and averaging (Kimoto et al., 2010). By acquiring the data as a series of sometimes several hundred low-dose frames over a period of what may be some minutes, the electron *dose-rate* is further lowered. Moreover, the intrinsically redundant nature of the data allows for unbiased correction of time-varying scanning-distortion/drift while preserving time-invariant sample data (Jones & Nellist, 2013; Sang & LeBeau, 2014; Jones et al., 2015; Ophus et al., 2016). These techniques also lend themselves to extension into areas including digital super-resolution (Bárcena-González et al., 2016, 2017) and simultaneous spectrum image recording and alignment (Yankovich et al., 2016; Wang et al., 2018; Jones et al., 2018) but discussion of these areas is left for a later work.

While appealing, reducing pixel dwell-time alone rapidly becomes insufficient as line flyback quickly dominates the electron exposure time. For conventional STEM scan rates, a relatively cautious flyback waiting time of 1000-1500 μs may be set as the default by the installing engineer (Sang et al., 2016). When this flyback time is reduced, appreciable distortions may become visible at the start of every line (typically the left half of the image), where the image appears compressed by up to 30 % (Buban et al., 2010). Fortunately, the profile of this artefact is highly reproducible and for times greater than around 10 μs appears to closely follow an exponential decay profile (Anderson et al., 2013; Kovarik et al., 2016; Buban et al., 2010; Sang et al., 2016); this suggests a route then to calibrate and correct for this in post-processing (Sanchez et al., 2006; Buban et al., 2010).

In this work, we explore the diagnosis and compensation of these flyback artefacts whose successful mitigation allows for greatly reduced flyback times to be used, delivering a commensurate increase in STEM frame-rate and reduction in electron exposure. First, we present a numerical study exploring the potential for dose-reduction. Second, methods for modelling and correcting the flyback compression artefact are discussed through comparison with generated, artefact-free reference frames. Finally, we experimentally apply our corrections to evaluate the approach and discuss the implications for fast multi-frame image averaging, and the practicalities of the real-world usage of this technique (Ophus et al., 2016; Sang & LeBeau, 2014).

## Background

*Electron-dose and Scanning Efficiency*

Before developing strategies to reduce electron dose, we will first revisit the current approach to dose calculation. The most common approach to calculate sample exposure requires the user to know their beam-current, and the spacing and dwell-time of each probe position. This leads to the expression in Equation 1:

$$Dose = I \cdot C\,(\delta_t)\, \frac{1}{(dx)^2} \qquad (1)$$

where, *I* is the beam-current in amps, *C* is the Coulomb number, $\delta_t$ is the dwell-time at each pixel in seconds and *dx* is the pixel width (assuming a square pixel), this equation calculates the dose in units of electrons per unit area. For CS acquisitions, where only a fraction of pixels are illuminated, this number is then multiplied by some fraction less than one. From this equation then, we see the motivation to pursue approaches with both low beam-currents and reduced dwell-times (Buban et al., 2010).

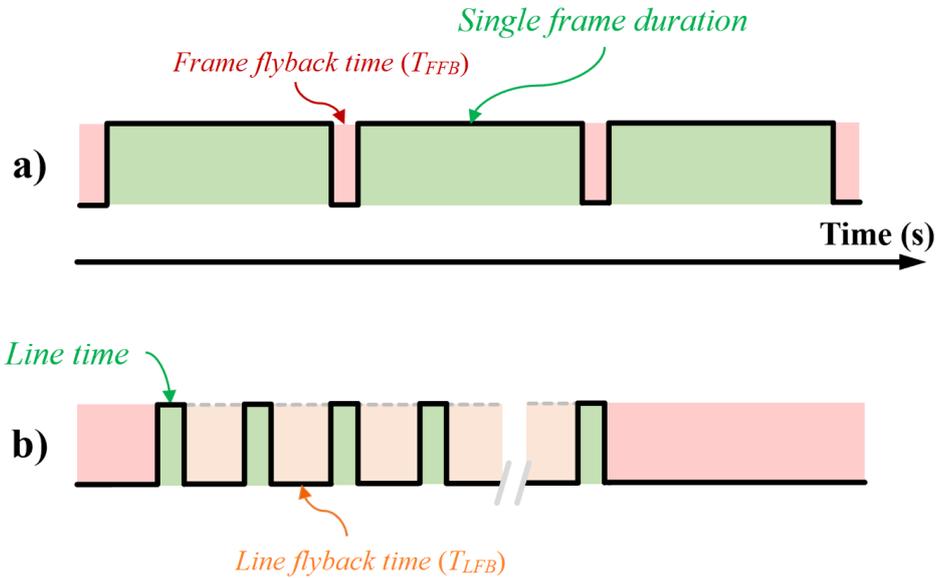

**Figure 1.** *Schematic illustration of sample electron-exposure with time. Plot a) shows an illustration of a series of 3 image frames recorded in succession over a characteristic time of several seconds. Plot b) shows an expanded illustration of the individual scan-lines within a single frame. Useful sample exposure which yields information is illustrated in green, while line flyback is shown in orange, and frame flyback time is shown in red.*

However, this equation represents a significant oversimplification of the practical operation of a STEM instrument. **Figure 1 (a)** shows a schematic representing the useful information-yielding electron exposure during a series of successive scanned image frames.

A conventional approach to producing images of high SNR is to capture a single frame with a very long dwell-time to allow the accumulation of a large amount of signal, however this leaves the data vulnerable to scanning noise and distortion (Muller et al., 2006). A more contemporary approach is to capture multiple fast image frames for later averaging, as this is more robust to these environmental effects (Sang & LeBeau, 2014; Jones et al., 2015). One might assume that an image derived from of 100 fast frames, and one of a single frame but with a 100 times longer dwell-time, expose the sample to equal amounts of electron radiation. However, this would only be true if the effects of two flyback times are ignored. These are the line-flyback time it takes for the electron beam to travel from the end of one image line to the beginning of the next, and the frame-flyback time, the time it takes for the beam to travel from the final pixel of an image frame to the first of the next.

For large dwell-times, these flyback times represent a small fraction of the overall frame time. However, **Figure 1 (b)** shows the detail *within* a single scan frame captured with a lower dwell-time and a characteristic time frame of milliseconds. In this scenario, not only is a significant amount of time expended during flyback time, but the electron probe still impinges on the sample. During this time, the probe may be just to the side of the area being image, but secondary electrons can still cause damage within the region of interest and beam-heating also persists (Egerton, 2017). These effects lead to potential sample damage while not delivering any useful information. Fast blanking during these times may be one solution , but again, this would require expensive hardware (Béché et al., 2016).

Knowing this motivated the creation of an expression for the efficiency with which our beam generates useful data, $\eta$:

$$\eta = \frac{\delta_t \cdot n_p}{(\delta_t \cdot n_p) + (T_{LFB} \cdot n_L) + T_{FFB}} \quad (2)$$

where $\delta_t$ is the pixel dwell-time, $n_p$ is the number of image pixels, $n_L$ is the number of scan-lines, $T_{LFB}$ is the line flyback time and $T_{FFB}$ is the frame flyback time. Equation 2 can be understood intuitively as the ratio of useful, information collecting time, to the total frame time (Mullarkey et al., 2020), and as $\eta$ is always less than 1 we can see that Equation 1 always underestimates the dose. Of the variables in this equation ($\delta_t$ , $n_p$ , $n_L$ , $T_{LFB}$, and $T_{FFB}$) generally only the first four out of five can be readily changed by the operator. Although increasing the pixel dwell-time and the number of pixels or decreasing the number of scan-lines would increase the efficiency, it would also increase the dose as per Equation (1). The route to increase scanning efficiency then, without increasing the dose, is to lower the line flyback time, if not eliminate it completely. This is of particular importance when using the aforementioned low-dose conditions of short dwell-times; the shorter each line is, the larger the fraction of it is occupied by the flyback time.

As an example, for a 512 × 512 pixel image captured with a 2 µs dwell-time, a 400 µs line flyback time and a 0.25 s frame flyback time, has an efficiency of just over 50 %. This means the dose predicted by equation (1) is approximately half of the actual dose the sample received. For a more general overview, we evaluate Equation 2 for dwell-times in the range 0.167 µs to 38 µs as this encompasses the entire range from the fastest Nion (6 MHz) and Gatan (2 MHz) scan-generators up to the line-sync speeds for 50 Hz mains power and 512 × 512 images. **Figure 2** shows the electron-dose efficiency across this range, frame flyback times were experimentally found on our system to be in the range of 0.2 s to 0.3 s and so 0.25 s was chosen as a representative value. Equation 2 was also evaluated for the same range of values for 256 × 256 images, resulting in overall lower efficiencies (**Supplementary Figure S1**).

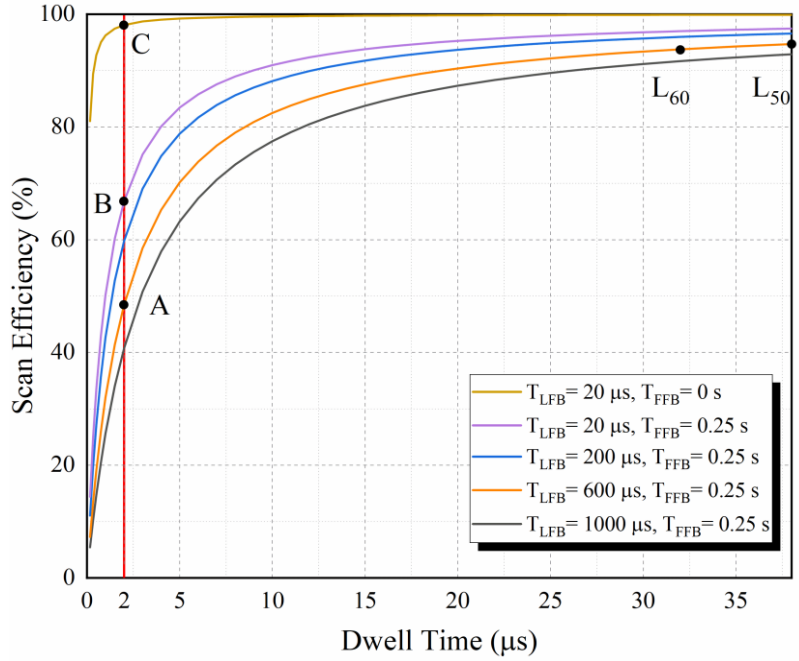

**Figure 2.** *Plots of dose-efficiency as a function of pixel dwell-time for varying flyback settings. The uppermost line represents the efficiencies which could be achieved with the elimination of frame flyback time. The points labelled $L_{50}$ and $L_{60}$ represent the efficiency of scans using a 600 µs line flyback time and which are line-synced to a 50 Hz and 60 Hz mains frequency respectively.*

At dwell-times around 38 µs, typical for traditional single frame imaging, the dose efficiency is above 90 % for all cases and perhaps explains why the wasted flyback dose has to date received little attention. For dwell-times of 2 µs (**Figure 2**, **red vertical line**), and where a 600 µs flyback time is maintained, the dose-efficiency falls to just under 50 %, indicating that more than half the total number of electrons passing through/near the image-region are causing damage but yielding no useful information. Alternatively, the dose efficiency improves to nearly 70 % for the same 2 µs dwell-time but for a flyback time of 20 µs, which we later show is usable without introducing distortions with the method introduced in this article. This jump in efficiency is represented by the points A and B in the figure and is strong motivation for using shorter flyback times. To jump again to point C requires the frame flyback time to be eliminated or reduced to a very low value. Approaches to achieve this may include custom scan generators or the implementation of a different pair of scan coils with low inductance, such as in (Ishikawa et al., 2020). These approaches are potentially costly or require dramatic modification to the instrument being used, which is generally not possible in a shared-user facility.

**Figure 2** also shows the case for line-synced captures at both 50 Hz and 60 Hz with a line flyback time of 600 µs, marked $L_{50}$ and $L_{60}$ respectively. In this case the start of each scan line is locked into the phase of the mains power supply, and when using a DigiScan the dwell-time is increased such that sequential scan-lines begin at the same phase. In both of these scenarios the efficiencies are very high at over 90 %, and the common use of line-syncing may explain why this topic has received little attention to date.

There have been approaches that seek to eliminate line flyback time entirely using either serpentine or spiral scans, but these approaches are somewhat complex to implement, require additional hardware, or result in circular images less convenient for onward processing (Wang et al., 2010; Sang et al., 2016). Here we focus on square (Shannon) scans available to all existing operators without the need to purchase new hardware. What remains then is to decrease the line flyback time as much as possible, but doing so begins to introduce artefacts into images due to the unavoidable hysteresis of the scan coils.

*Scan Coil Hysteresis*

As the scan-coils are electromagnets of finite inductance there is a non-zero response time of the coil current to changes in the driving voltage (Anderson et al., 2013). Due to this there is an appreciable lag between the expected position of the electron beam and its achieved position until the beam reaches a constant velocity (Velazco et al., 2020). This is analogous to inertia in a physical system, and leads to distortions in images captured at low flyback times (Sang et al., 2016; Buban et al., 2010).

These distortions can be understood by considering the physical path the beam tracks during the flyback time. After reaching the end of a scan-line the beam initially travels from right-to-left (for a typical scan) and overshoots for first pixel of the next line. It must then travel from left-to-right, ready for the next line's start. At long line flyback times the beam has enough time to reach a constant velocity as it is travelling to the right to begin the next line. However, with short flyback times the beam is still lagging behind the correct position when the next scan-line begins. This causes an overly large area of the sample to be captured and fitting this extra area into the image pixels results in the appearance of compressed data at the start of each line. Although this may vary by instrument, this is indeed observed in this article on two different microscopes (**Supplementary Figure S2**). Conventionally the electron-dose is highest at the left-hand side of an image due to the flyback time. However, at the conditions used here where the beam is travelling faster at the start of each line, the electron dose may be lower in this region as a larger area is scanned in the same number of pixel dwell times, as evidenced by the compression artefact.

It has been observed that the scan-coil rise time follows a characteristic exponential shape, as to be expected from a system with inductance effects (Buban et al., 2010; Anderson et al., 2013). In extreme cases where the flyback time is eliminated, and direct or indirect measures of the deflection rise-time can be measured, 90 % rise-time values between 10-100 µs have been obtained (Anderson et al., 2013; Kovarik et al., 2016). At typical scanning speeds this rise-time would be confined in only the first few pixels of a line. However at shorter dwell-times, this artefact becomes more apparent and may cover half the width of the frame (Buban et al., 2010; Sang et al., 2016).

## Methods

The experimental results presented here were recorded using a Nion UltraSTEM 200 at 200 kV controlled through a Gatan DigiScan II, unless stated otherwise. Similar, though not identical, effects were seen on our ThermoFisher Titan STEM with its own PIA scan-generator, but for clarity we focus on results from the common DigiScan controlled system.

*Automated Flyback Diagnosis*

Various approaches may be followed to measure the flyback distortion when using a single-crystal reference sample. For example geometric-phase analysis (GPA) can be used to diagnose distortions, where the $e_{xx}$ term from a perfect single-crystal image natively captures flyback distortion (Hÿtch et al., 1998; Zhu et al., 2013). However, owing to Fourier truncation, this technique always exhibits some edge effects around the image in the measured distortion fields (unfortunately the region of most interest). Moreover, depending on the radius, x-y position, and edge-smoothing of the manually selected Fourier masks (minimum 2 masks leads to a minimum of 6 adjustable parameters), other systematic artefacts may be introduced (Peters et al., 2015).

Alternatively, a manual analysis extracting the position of fiducial features (Anderson et al., 2013), or the relative compression of crystal unit-cells (Buban et al., 2010) may be performed. However, this again requires manual intervention and/or special sample orientation.

What is ideally needed then, for this to be readily deployable to a general STEM user, is some fully automated method (e.g. without manual selection of Fourier masks), and which does not intrinsically have additional image-edge artefacts. The solution to both these problems is to work in real-space (rather than reciprocal-space) and use a readily available single-crystal reference material (e.g., Si, $SrTiO_3$). The regular atomic spacings of such a material can be used to optimise any parameters to define the exponential decay.

Using the prior knowledge that the flyback hysteresis causes each scan-line to be compressed, which can be approximated with an exponential form, we can then write the distortion as a function of just two unknown parameters (Buban et al., 2010):

$$x_{corrected} = x - Ae^{-t/b} \quad (3)$$

where $x$ are the measured x-positions and $A$ and $b$ are to be determined. Calculating the exponential compression and subtracting it from the measured positions will correct this compression artefact, restoring the true positions, $x_{corrected}$ (equation (3)). Working with units of time in the exponential has the advantage that we can deal explicitly with the flyback time:

$$t = x \cdot \delta_t + T_{LFB} \quad (4)$$

This simple distortion then is a function described by two parameters, $A$ and $b$, and can be removed from an image using bilinear interpolation. The instrument-specific parameters can be obtained from non-linear least squared fitting with respect to a distortion free reference image, sometimes referred to as a quasi-static reference image (Rečnik et al., 2005; Sanchez et al., 2006). This reference can be generated in two ways. Either an image is acquired with large flyback (e.g. during calibration when the microscope is first installed) or we can use the fact that the furthest right half of an image will have negligible distortion due to flyback (**Supplementary Figure S3**). In either case, the reference needs to be made with the same crystallographic phase as the image to be diagnosed. The method here uses autocorrelation to perform self-same tiling and generate an oversized image, from which an appropriately positioned reference can be extracted (**Supplementary Figure S4**).

The fitting process can be repeated for images of varying line flyback and dwell-time combinations with the calculated values for *A* and *b* tabulated for each. With the observation that these parameters remain constant for the same imaging settings at the same operating voltages, even across a timespan of weeks (**Supplementary Figure S5**), these values can then be used to correct any future images taken at the same settings.

An expansion of this, is to generalise the recorded values of *A* and *b* and to use these values to generate an empirical equation to describe the behaviour of the scan coils. This instrument-specific equation can then be used to determine the correct values of *A* and *b* which eliminate the hysteresis artefact in an image for all combinations of flyback-time and dwell-time. This is the approach taken in this article as a given instrument can be characterised (say by core staff) and then be available to all users. It should be noted that, in the same way that magnification is calibrated on a per-instrument basis, so too this inductance descriptor equation should be done on a per-instrument basis.

The model here assumes a single time constant *b* for the system. Although this is unlikely to be the case due to different elements of the scan system such as the waveform generator, filter boards, and coil magnetisation each having their own time constants, it is seen here that this approximation leads to satisfactory results for a practical range of line flyback and pixel dwell times.

*Measuring $T_{FFB}$*

Although the line flyback time can be set by the user using a DigiScan II, the frame flyback time, $T_{FFB}$, cannot. This is the time taken for the electron beam to travel from the final pixel of an image frame to the first pixel of the next. To measure this, an acquisition was run for 60 seconds with known imaging conditions. Assuming a $T_{FFB}$ of zero, the frametime, and therefore expected number of frames in one minute can be calculated. This can then be compared to the experimentally achieved number of frames, with the difference being explained by any unaccounted-for time, in this case the frame flyback time.

*Samples*

The samples used in this article are lamellae of strontium titanate ($SrTiO_3$) and silicon. These were chosen as they are readily available single crystals whose real-space periodicity is exploited in our method as described above. Any sample can be used if a long flyback quasi-static image is chosen as a reference, however this requires minimal sample drift for the duration of the data-capture for best performance. In practice a sample with periodic features that can be tiled, as shown in **Supplementary Figure** S4, is more suitable and is recommended.

## Results & Discussion

Averaging a series of frames of an object is a common means of improving image SNR. Recently, more advanced tools for correcting for non-rigid distortions before this averaging have been shown to improve performance further (Berkels et al., 2014; Jones et al., 2015; Ophus et al., 2016). The algorithms in these previous studies work in slightly different ways, but all rely on one assumption; that the multiple observations provide a mathematically redundant set from which the time-varying distortions can be separated from the time-invariant ground-truth. Additional redundancy can be encoded into the dataset by varying the scan-orientation during the acquisition, either as perpendicular pairs (Ophus et al., 2016) or with images recorded with a 90° increment between successive frames (Sang & LeBeau, 2014; Jones et al., 2015). Data from both single-orientation series and rotating-scan frame series are presented in our results, with the former presented first.

*Hysteresis Correction of a Single-orientation Series*

For a qualitative visual comparison excerpts from images of strontium titanate (STO) with increasingly shorter flyback waiting times are shown in **Figure 3 (a)**.

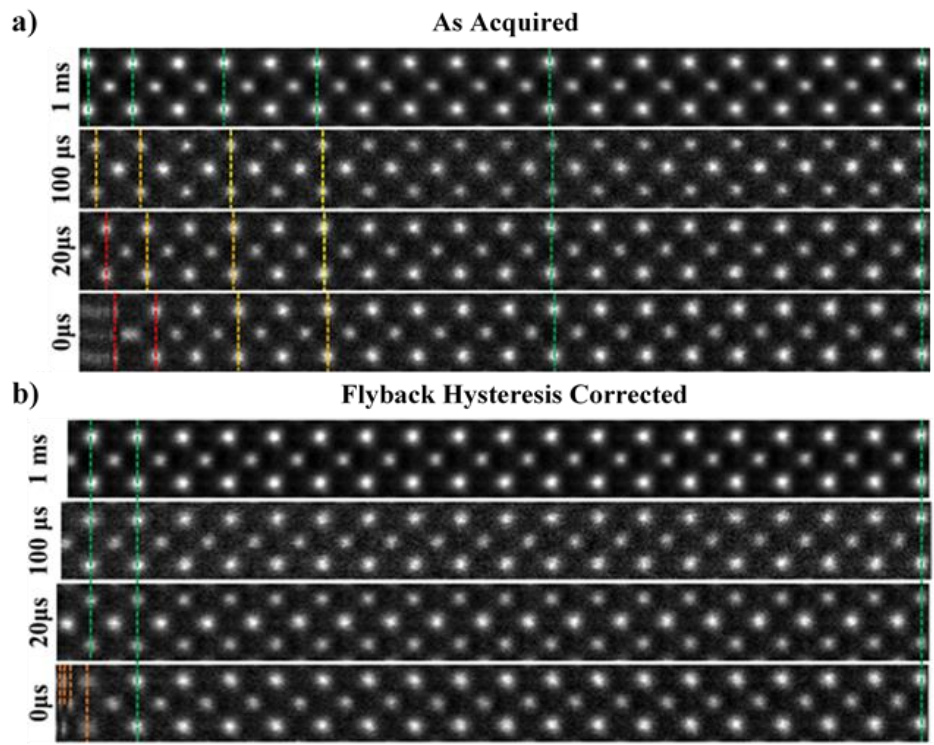

**Figure 3, Top**: *Montage of cropped panels from increasing faster line flyback time acquisitions. At faster acquisitions the unit-cell appears increasingly compressed.* **Bottom:** *A montage of the same images post-correction. For flyback times as low as 20 µs visually fidelity is retained across the entire field of view. Some artefact remains in the leftmost region of the 0 µs image.*

The image captured with the 1000 µs (1 ms) line flyback time is taken to be a quasi-static reference frame for visual comparison with the distorted images. Across the entire width of the quasi-static frame the unit-cells are uniform in width (**Supplementary Figure S3**). Equally, at all flyback-delays the latter half of each scan-line (right-hand half) is also in alignment with the 1 ms frame, which is good justification for using this region to generate a reference image to compare against.

At shorter flyback-delays the leftmost region of the image is increasingly distorted (compressed). Using the determined hysteresis parameters, the flyback distortion is corrected in all raw-frames prior to any alignment and non-rigid registration. **Figure 3 (b)** shows the same data to **Figure 3 (a)** but after flyback correction.

As shown in **Supplementary Figure S6** this is reduced to below 0.1 % After applying this correction these distortions are eliminated for flyback times as low as 20 μs. In the case of 0 μs line flyback time only a very small region at the beginning of the line still retains some distortion. This is because the beam cannot achieve a 0 μs flyback time as it takes a finite amount of time. The remaining distortion then is what is captured while the beam is still flying back.

For a quantitative comparison the diagnosis of the compression using the proposed automated method and also manually following the method in (Buban et al., 2010), before and after correction are presented in **Figure 4**.

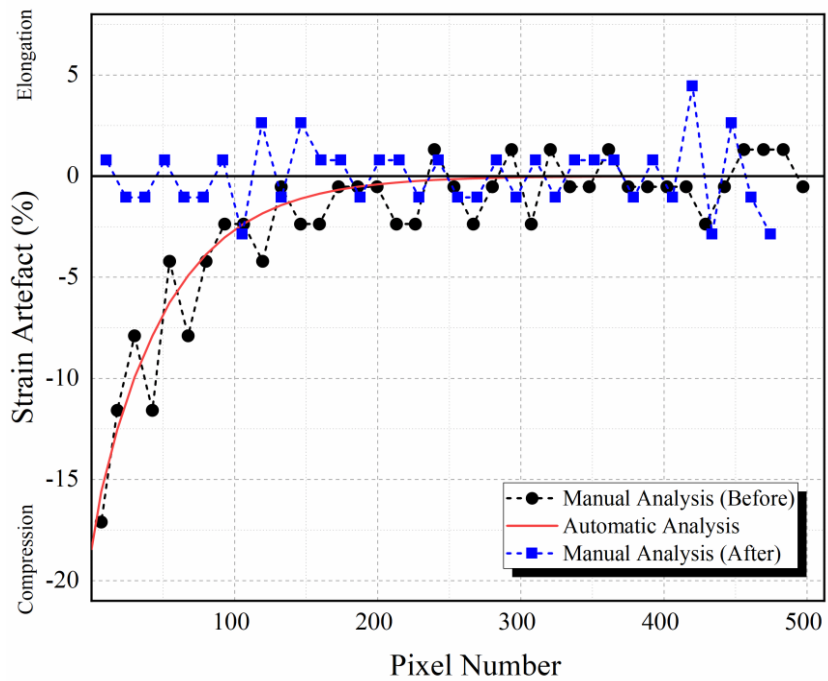

**Figure 4.** *Example flyback hysteresis diagnosis for a 512 px wide image with a 20 μs flyback and 0.5 μs dwell-time. The black round points represent a manual analysis after Buban et al.(Buban et al., 2010), while the solid line represents the output of our automated analysis method introduced here. The blue square points represent a manual analysis after flyback hysteresis correction.*

**Figure 4** shows an example analysis for the 20 μs flyback-delay acquisition with both the results of the automated parametric diagnosis and the manual unit-cell based analysis. While both methods give nominally identical results, the manual method is limited in its precision because of human intervention and a small random scatter due to the finite pixilation of the peak position identification. The automated approach however uses a smoothly varying mesh as opposed to pixels, uses the entire field-of-view, and does not rely on any particular scan orientation. Due to this the automated approach is used going forward.

*Diagnosing the Hysteresis Compression Artefact*

The method described above was used to then diagnose the compression for a wide range of dwell-times and line flyback times with fixed scan rotation angle. **Figure 5** shows the experimentally determined *A* and *b* values for a range of dwell times and line flyback times. Clear trends with dwell time are present for both parameters, but noticeable the use of equation (4) has removed any flyback dependence, allowing the diagnosis to depend only on the dwell-time and not the line flyback time.

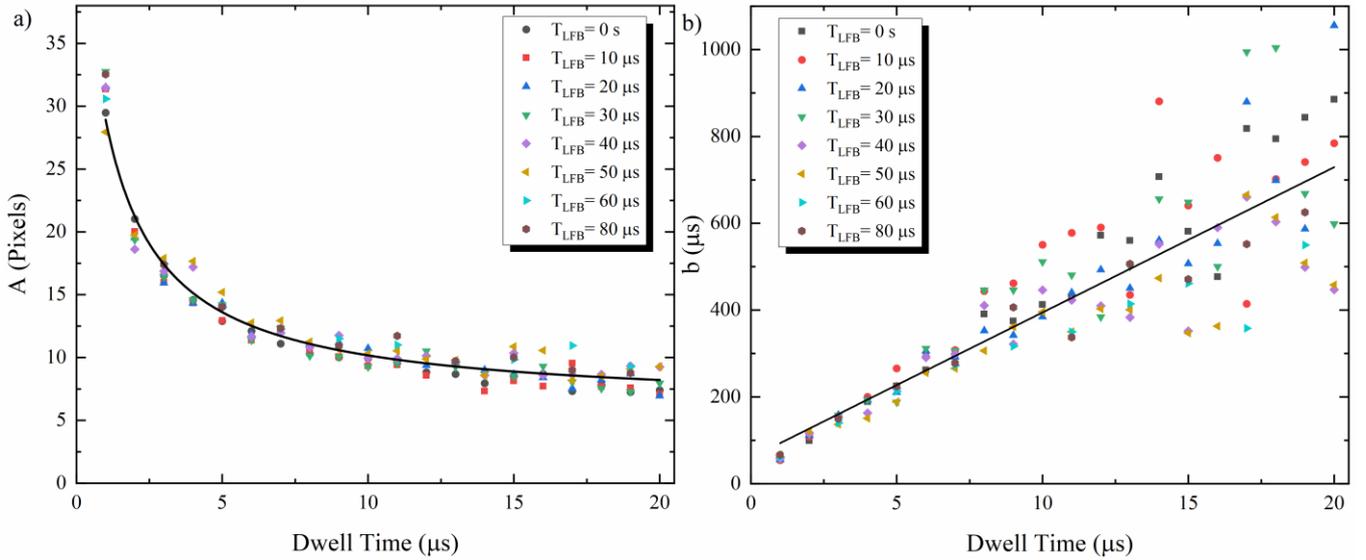

**Figure 5.** *Graphs of A and b from equation (3) which describes the compression of each scan line due to flyback hysteresis. These are presented as a function of dwell time for varying line flyback times. Error estimates for the fit parameters and goodness-of-fit estimates can be found Table S1 in the Supplementary Information.*

The *A* parameter in **Figure 5 a)** clearly shows an inverse decay with dwell time whilst **Figure 5 b)** shows a linear increase of *b* as the dwell-time. This is understandable, as the slower scans exhibit a reduced magnitude of compression (lower *A* value), and the longer dwell-times exhibit a more rapid decay of the compression across the image (larger *b* value).

From this we can see that a relatively simple set of equations (one linear, one inverse) can be used to determine the parameters needed in equation (3) to correct flyback distortions for all line flybacks and dwell times. As the equations do not depend on the flyback time this means only a small set of images is required and with only dwell time varying to fully diagnose the hysteresis behaviour. Discussion on the scatter of the points in **Figure 5 b)** is included in the caption of **Supplementary Figure S5**, while the mean and standard deviation of the data are plotted in **Supplementary Figure S7**.

The physical origin of these functions is obfuscated by each manufacturer's choice of the microscope control hardware and the scan generator itself. For example, the DigiScan II produces a pure sawtooth waveform, even during line flyback wait time, that modifies the coil current of the left side of the image dependent on the scan speed (and therefore dwell time). To avoid unnecessary discussion that is specific to our particular microscope, and to maintain generality, we do not attempt to reduce the number of fit parameters further. Nevertheless, though we might expect the exact form of the trends to vary between microscopes, the observed trends are consistent can be used to diagnose and correct the compression.

Due to the limited range of fields-of-view which can be used on the Nion UltraSTEM 200 a full characterisation of the effects of magnification on this process could not be carried out. However, magnification is linearly related to the current in the scan coils, and we predict that the effect of the visible hysteresis within the image will also scale linearly. We therefore expect that fractionally, the compression artefact will appear the same within any image at different magnifications. However, when changing from very high to very low magnifications the current in the scan coils can change by an order of magnitude and this assumption may not hold, and so we recommend operators investigate this on their own instruments where necessary. Similarly, as most operators choose between a small set of fixed voltages we do not fit the voltage dependence and instead, for simplicity, recommending characterising each voltage separately.

*Correction of Rotating Scan Frame-series*

Where flyback hysteresis is present in each frame of a scan-rotated series, one edge will be corrupted for each of the principal scan-orientations; this has significant effects for both rigid and non-rigid registration. For rigid registration (correction of simple stage drift translation) the differing artefact profiles in each scan-orientation reduces the quality of the cross-correlation (or other correlation) used to align the data and may lead to poorer drift diagnosis. For non-rigid registration algorithms, where the average frame is used as a starting estimate (Jones et al., 2015), the degraded starting average frame may reduce the quality and speed of the registration convergence. Instead, by pre-treating the multi-frame data to first correct for the now calibrated flyback hysteresis, the speed and precision of the non-rigid correction can be improved. After pre-treatment, the fast-flyback multi-frame series, can be processed as if a conventional series and corrected for rigid-offsets (Bárcena-González et al., 2016), affine-shears (Dycus et al., 2015; Sang & LeBeau, 2014), and non-linear distortions (Jones et al., 2015; Ophus et al., 2016; Wang et al., 2018) as necessary.

A series of 100 images of $SrTiO_3$, each with a 20 μs line flyback time and a scan rotation of 90° between consecutive frames was captured to demonstrate the ability of this technique to correct flyback hysteresis artefacts for rotating image series. This image series was aligned and then averaged with, and without, hysteresis correction applied as the first step, with the results shown in **Figure 6**. The image comparison in **Figure 6** shows the same image series both with and without flyback hysteresis correction applied as the first post-processing step before the usual aligning and averaging. While the flyback-uncorrected average image (**Figure 6 (a)**) shows the distortions on all edges, the flyback-pretreated average image (**Figure 6 (b)**) maintains fidelity across the entire field of view.

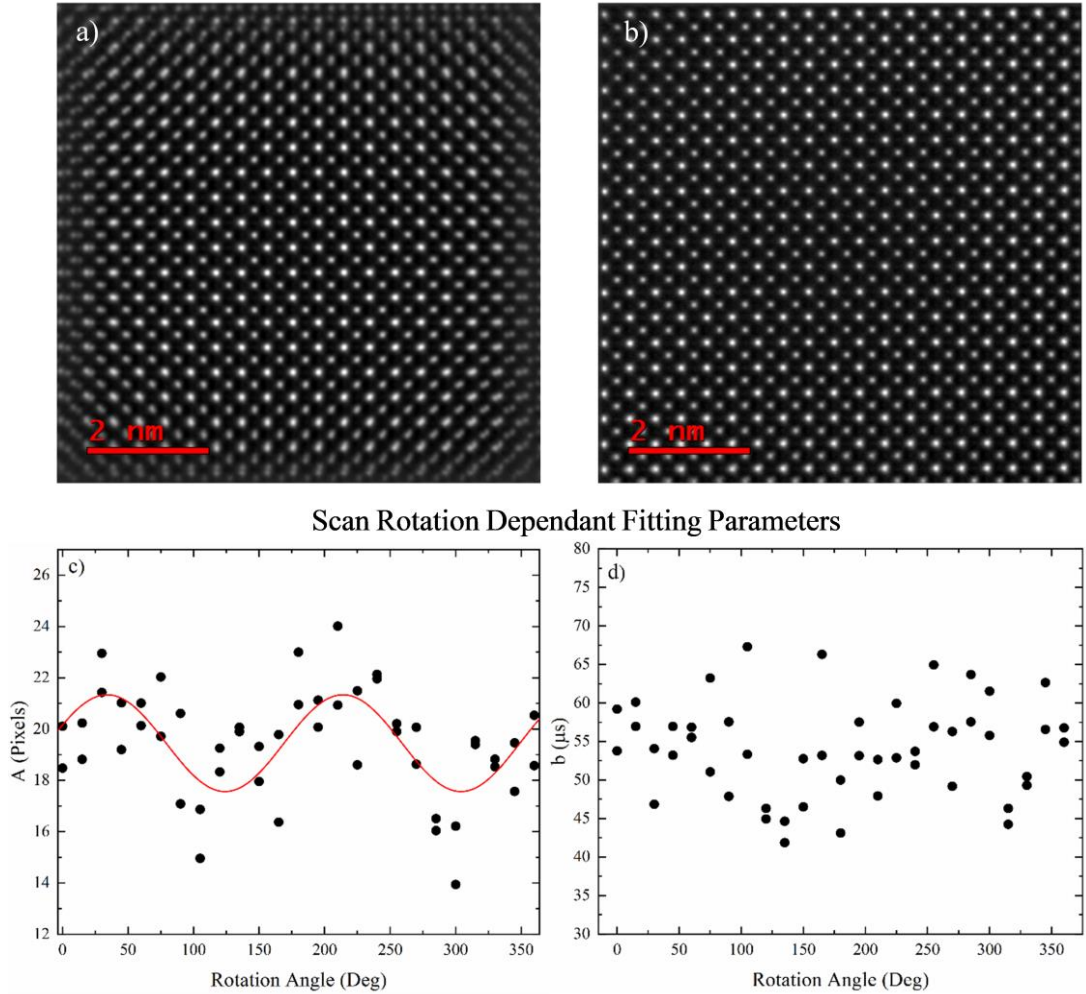

**Figure 6.** *Aligned average of 100 ADF frames recorded with a 20 µs flyback time and a scan-rotation increment of 90° between frames. Average frame is shown without hysteresis correction (**a**) and with the proposed correction (**b**). (**c**), (**d**) show the variation of A and b, respectively, with rotation angle. The images analysed were captured with a dwell-time of 2 µs and line flyback time of both 0 s and 20 µs with the rotation angle increasing in steps of 15° from 0° to 360°.*

Averaging multiple frames is an important approach to increase the SNR where it is too low in a single frame, and also to encode redundancies into our datasets to aid in future processing. One reason the SNR can be too low in a single frame is when low dwell-times are being used, which is the region where the largest jumps in scanning efficiency are seen when the line flyback time is reduced (**Figure 2**). It is therefore of key importance that the method presented here is compatible with multiframe imaging, as is shown.

**Figure 6 (b)** was produced by applying hysteresis correction as the first step to each captured image frame individually. Although this works on a case-by-case basis, if we wish to construct a comprehensive model of the hysteresis, the general effects of scan rotation on the compression artefact should be investigated. To do so, a series of images of varying dwell times and line-flyback times were captured while *also* varying the scan rotation angle through a full 360° (a total of 50 images were recorded). The hysteresis diagnosis step was performed for each, with the calculated values for *A* and *b* plotted as a function of rotation angle in **Figure 6** (**c,d**).

**Figure 6 (c,d)** shows that while *b* does not appear to vary in any meaningful way with the rotation angle, *A* varies sinusoidally with rotation angle. The variation could be explained by factors such as different coil windings in the x and y scan-coils. Another feature to note is that the cosine function is not at a maximum at a rotation angle of zero, i.e., there is an offset. This was allowed as a parameter in the fit and could be explained by a physical assembly misalignment or a manufacturer firmware offset (e.g. for the orientation of the stage to the Ronchigram camera). Although the exact physical origins of these effects are not exactly known, reassuringly it can be seen that there are two cycles of the sinusoid in a full 360° rotation, corresponding to the two-fold rotation symmetry of the scan coils.

Having established the variations of *A* and *b* with rotation angle, dwell-time, and line-flyback time we now have all the empirical coefficients to fully describe the flyback hysteresis effects in our microscope. Simply knowing these three imaging parameters will allow the automatic correction of *any* future image captured and could even be applied retroactively to any previously captured data.

## Conclusions

In this paper we introduce the concept of scanning efficiency and lay bare the unfortunate reality that it can be surprisingly low for increasingly commonplace fast-frame imaging conditions. To increase the efficiency without also increasing the electron dose, the path to take is to decrease the line-flyback time as much as is possible. Unfortunately, at low flyback times a compression artefact due to hysteresis in the scan coils appears.

We have demonstrated the simple characterisation of flyback distortions and the subsequent correction for this on single image frames, and multi-frame datasets with or without scan rotation on subsequent frames. Not only is precision retained across the full field-of-view, ensuring techniques such as strain-mapping remains precise even when using low flyback times, but the scanning efficiency is increased (**Supplementary Figure S6**). This paper describes the jump in efficiency from 48 % (**Figure 2, point 'A'**) to 67 % (**Figure 2**, **point 'B'**), an increase of almost 20 %. This figure also indicates that to achieve very high efficiencies, such as labelled 'C', we need a separate approach to control the frame flyback time, such as a custom scan generator.

Furthermore, we have shown that it is possible to form a semi-empirical model of the hysteresis effect to be created, whereby any image captured with any arbitrary combination of line-flyback time, dwell-time, and scan rotation can be corrected. Although this model is expected to vary by instrument, the method presented in this manuscript can, and should be, repeated by operators for their own instruments. This will enable the entire community of STEM operators to reduce electron-dose, sample-damage, and increase frame-rates without any additional hardware purchase.

## Code & Data Availability

The code which is used to measure and correct flyback distortions can be found at our Github repository https://github.com/TCD-Ultramicroscopy/STEM-Flyback-correction . The data used in this paper can be found at doi.org/ 10.5281/zenodo.5713336.

## Acknowledgments

The authors would like to acknowledge the Centre for Research on Adaptive Nanostructures and Nanodevices (CRANN) and the Advanced Materials and BioEngineering Research (AMBER) Network for financial and infrastructural support for this work. J.J.P.P. and L.J. acknowledge SFI grant 19/FFP/6813, T.M. acknowledges the SFI & EPSRC Centre for Doctoral Training in the Advanced Characterisation of Materials (award references 18/EPSRC-CDT-3581 and EP/S023259/1).

**Competing Interests**: The authors declare no competing interests.

# References

ANDERSON, H. S., ILIC-HELMS, J., ROHRER, B., WHEELER, J. & LARSON, K. (2013). Sparse imaging for fast electron microscopy. *IS&T/SPIE Electronic Imaging* **8657**, 86570C.

BÁRCENA-GONZÁLEZ, G., GUERRERO-LEBRERO, M. P., GUERRERO, E., FERNÁNDEZ-REYES, D., GONZÁLEZ, D., MAYORAL, A., UTRILLA, A. D., ULLOA, J. M. & GALINDO, P. L. (2016). Strain mapping accuracy improvement using super-resolution techniques. *Journal of Microscopy* **262**, 50–58.

BÁRCENA-GONZÁLEZ, G., GUERRERO-LEBRERO, M. P., GUERRERO, E., YAÑEZ, A., FERNÁNDEZ-REYES, D., GONZÁLEZ, D. & GALINDO, P. L. (2017). Evaluation of high-quality image reconstruction techniques applied to high-resolution Z-contrast imaging. *Ultramicroscopy* **182**, 283–291.

BÉCHÉ, A., GORIS, B., FREITAG, B. & VERBEECK, J. (2016). Development of a fast electromagnetic beam blanker for compressed sensing in scanning transmission electron microscopy. *Applied Physics Letters* **108**, 093103.

BERKELS, B., BINEV, P., BLOM, D. A., DAHMEN, W., SHARPLEY, R. C. & VOGT, T. (2014). Optimized imaging using non-rigid registration. *Ultramicroscopy* **138**, 45–56.

VAN DEN BROEK, W., REED, B. W., BECHE, A., VELAZCO, A., VERBEECK, J. & KOCH, C. T. (2019). Various Compressed Sensing Setups Evaluated Against Shannon Sampling Under Constraint of Constant Illumination. *IEEE Transactions on Computational Imaging* **5**, 502–514.

BUBAN, J. P., RAMASSE, Q., GIPSON, B., BROWNING, N. D. & STAHLBERG, H. (2010). High-resolution low-dose scanning transmission electron microscopy. *Journal of Electron Microscopy* **59**, 103–112.

CHEN, Q., DWYER, C., SHENG, G., ZHU, C., LI, X., ZHENG, C. & ZHU, Y. (2020). Imaging Beam-Sensitive Materials by Electron Microscopy. *Advanced Materials* **32**, 1907619.

DYCUS, J. H., HARRIS, J. S., SANG, X., FANCHER, C. M., FINDLAY, S. D., ONI, A. A., CHAN, T. E., KOCH, C. C., JONES, J. L., ALLEN, L. J., IRVING, D. L. & LEBEAU, J. M. (2015). Accurate Nanoscale Crystallography in Real-Space Using Scanning Transmission Electron Microscopy. *Microscopy and Microanalysis* **21**, 946–952.

EGERTON, R. F. (2017). Scattering delocalization and radiation damage in STEM-EELS. *Ultramicroscopy* **180**, 115–124.

ELBAUM, M. (2018). Quantitative Cryo-Scanning Transmission Electron Microscopy of Biological Materials. *Advanced Materials* **30**, 1706681.

GLAESER, R. M. (1971). Limitations to significant information in biological electron microscopy as a result of radiation damage. *Journal of Ultrasructure Research* **36**, 466–482.

HŸTCH, M. J., SNOECK, E. & KILAAS, R. (1998). Quantitative measurement of displacement and strain fields from HREM micrographs. *Ultramicroscopy* **74**, 131–146.

ISHIKAWA, R., JIMBO, Y., TERAO, M., NISHIKAWA, M., UENO, Y., MORISHITA, S., MUKAI, M., SHIBATA, N. & IKUHARA, Y. (2020). High spatiotemporal-resolution imaging in the scanning transmission electron microscope. *Microscopy* **69**, 240–247.

JONES, L. & DOWNING, C. (2018). The MTF & DQE of Annular Dark Field STEM: Implications for Low-dose Imaging and Compressed Sensing. *Microscopy and Microanalysis* **24**, 478–479.

JONES, L. & NELLIST, P. D. (2013). Identifying and correcting scan noise and drift in the scanning transmission electron microscope. *Microscopy and Microanalysis* **19**, 1050–1060.

JONES, L., VARAMBHIA, A., BEANLAND, R., KEPAPTSOGLOU, D., GRIFFITHS, I., ISHIZUKA, A., AZOUGH, F., FREER, R., ISHIZUKA, K., CHERNS, D., RAMASSE, Q. M., LOZANO-PEREZ, S. & NELLIST, P. D. (2018). Managing dose-, damage- and data-rates in multi-frame spectrum-imaging. *Microscopy* **67**, 98–113.

JONES, L., YANG, H., PENNYCOOK, T. J., MARSHALL, M. S. J., VAN AERT, S., BROWNING, N. D., CASTELL, M. R. & NELLIST, P. D. (2015). Smart Align—a new tool for robust non-rigid registration of scanning microscope data. *Advanced Structural and Chemical Imaging* **1**, 1–16.

KIMOTO, K., ASAKA, T., YU, X., NAGAI, T., MATSUI, Y. & ISHIZUKA, K. (2010). Local crystal structure analysis with several picometer precision using scanning transmission electron microscopy. *Ultramicroscopy* **110**, 778–782.

KOVARIK, L., STEVENS, A., LIYU, A. & BROWNING, N. D. (2016). Implementing an accurate and rapid sparse sampling approach for low-dose atomic resolution STEM imaging. *Applied Physics Letters* **109**, 164102.

MITTELBERGER, A., KRAMBERGER, C. & MEYER, J. C. (2018). Software electron counting for low-dose scanning transmission electron microscopy. *Ultramicroscopy* **188**, 1–7.

MULLARKEY, T., DOWNING, C. & JONES, L. (2020). Development of a Practicable Digital Pulse Read-Out for Dark-Field STEM. *Microscopy and Microanalysis* **27**, 99–108.

MULLER, D. A., KIRKLAND, E. J., THOMAS, M. G., GRAZUL, J. L., FITTING, L. & WEYLAND, M. (2006). Room design for high-performance electron microscopy. *Ultramicroscopy* **106**, 1033–1040.

OPHUS, C., CISTON, J. & NELSON, C. T. (2016). Correcting nonlinear drift distortion of scanning probe and scanning transmission electron microscopies from image pairs with orthogonal scan directions. *Ultramicroscopy* **162**, 1–9.

PETERS, J. J. P., BEANLAND, R., ALEXE, M., COCKBURN, J. W., REVIN, D. G., ZHANG, S. Y. & SANCHEZ, A. M. (2015). Artefacts in geometric phase analysis of compound materials. *Ultramicroscopy* **157**, 91–97.

REČNIK, A., MÖBUS, G. & ŠTURM, S. (2005). IMAGE-WARP: A real-space restoration method for high-resolution STEM images using quantitative HRTEM analysis. *Ultramicroscopy* **103**, 285–301.

SANCHEZ, A. M., GALINDO, P. L., KRET, S., FALKE, M., BEANLAND, R. & GOODHEW, P. J. (2006). An approach to the systematic distortion correction in aberration-corrected HAADF images. *Journal of Microscopy* **221**, 1–7.

SANDERS, T. & DWYER, C. (2018). Inpainting Versus Denoising for Dose Reduction in STEM. *Microscopy and Microanalysis* **24**, 482–483.

SANG, X. & LEBEAU, J. M. (2014). Revolving scanning transmission electron microscopy: Correcting sample drift distortion without prior knowledge. *Ultramicroscopy* **138**, 28–35.

SANG, X., LUPINI, A. R., UNOCIC, R. R., CHI, M., BORISEVICH, A. Y., KALININ, S. V., ENDEVE, E., ARCHIBALD, R. K. & JESSE, S. (2016). Dynamic scan control in STEM: spiral scans. *Advanced Structural and Chemical Imaging* **2**, 6.

STEVENS, A., YANG, H., CARIN, L., ARSLAN, I. & BROWNING, N. D. (2014). The potential for Bayesian compressive sensing to significantly reduce electron dose in high-resolution STEM images. *Microscopy* **63**, 41–51.

STEVENS, A., YANG, H., HAO, W., JONES, L., OPHUS, C., NELLIST, P. D. & BROWNING, N. D. (2018). Subsampled STEM-ptychography. *Applied Physics Letters* **113**, 033104.

VELAZCO, A., NORD, M., BÉCHÉ, A. & VERBEECK, J. (2020). Evaluation of different rectangular scan strategies for STEM imaging. *Ultramicroscopy* **215**, 113021.

WANG, JUNTING, WANG, JIHUI, HOU, Y. & LU, Q. (2010). Self-manifestation and universal correction of image distortion in scanning tunneling microscopy with spiral scan. *Review of Scientific Instruments* **81**, 073705.

WANG, Y., EREN SUYOLCU, Y., SALZBERGER, U., HAHN, K., SROT, V., SIGLE, W. & VAN AKEN, P. A. (2018). Correcting the linear and nonlinear distortions for atomically resolved STEM spectrum and diffraction imaging. *Microscopy* **67**, i114–i122.

YANKOVICH, A. B., BERKELS, B., DAHMEN, W., BINEV, P. & VOYLES, P. M. (2015). High-precision scanning transmission electron microscopy at coarse pixel sampling for reduced electron dose. *Advanced Structural and Chemical Imaging* **1**, 2.

YANKOVICH, A. B., ZHANG, C., OH, A., SLATER, T. J. A., AZOUGH, F., FREER, R., HAIGH, S. J., WILLETT, R. & VOYLES, P. M. (2016). Non-rigid registration and non-local principle component analysis to improve electron microscopy spectrum images. *Nanotechnology* **27**, 364001.

ZHU, Y., OPHUS, C., CISTON, J. & WANG, H. (2013). Interface lattice displacement measurement to 1 pm by geometric phase analysis on aberration-corrected HAADF STEM images. *Acta Materialia* **61**, 5645–5663.

# Supplementary materials for Using Your Beam Efficiently: Reducing Electron-dose in the STEM via Flyback Compensation

Tiarnan Mullarkey[1,2], Jonathan J. P. Peters[1], Clive Downing[3], Lewys Jones[1,3]

1. School of Physics, Trinity College Dublin, Dublin 2, Ireland.
2. Centre for Doctoral Training in the Advanced Characterisation of Materials, AMBER Centre, Dublin 2, Ireland.
3. Advanced Microscopy Laboratory, Centre for Research on Adaptive Nanostructures & Nanodevices (CRANN), Dublin 2, Ireland.

## Reference generation

In order to determine the parameters used for correcting the flyback distortion, comparison with an undistorted reference is required. We propose two choices for a reference image: an image with large flyback time (e.g. 1 ms) or the right half of the image with flyback hysteresis effects. **Figure S3 (a)** and **(b)** show images of silicon <110> with 1 ms and 20 µs flyback, respectively. The distortion in the direction of the scanlines can be measured using geometric phase analysis (GPA) (Hÿtch et al., 1998; Peters, 2021), shown in **Figs. S3 (c)** and **(d)**, with an expected zero strain from bulk silicon. As expected, the 1 ms flyback shows zero strain across the entire image and the 20 µs flyback image shows a distortion on the left side. Note that both images exhibit the edge effect artefacts due to the periodic nature of the Fourier transforms used in GPA. However, it can be seen that the left half of the image is relatively flat and with no distortion. This is highlighted further in the average strain profiles shown in **Figs. S3 (e)** and **(f)**.

As dwell times are reduced, it may be expected that the distortion free area may reduce and become unsuitable for use as a reference. However, within the capabilities of the Gatan Digiscan II and Nion UltraSTEM, we did not find this to be the case. Use of a reference from the same image that need to be corrected does however have some advantages, particularly when considering stability of the imaging system. Any aberrations within the image will also be present in the reference as well as any constant drift effects (i.e. as given by an affine transformation). Similarly, for short dwell time images there is often a poor signal to noise ratio and frame averaging may need to be used. As the use of registration will alter the perceived hysteresis effects, direct summing must be used and will lead to image blurring for any amount of stage drift. These effects reduce error in the least squares fitting calculation and can increase the quality of the fit. However, for both choices of reference image, the crystallography needs to be aligned with the image to be corrected.

The generation of an aligned reference is performed by generating an oversized reference from self-tiling (similar to the template matching algorithms in Smart Align (Jones et al., 2015)). Here we take the choice of undistorted image and use cross correlation to find two non-colinear, non-zero vectors that map the image onto itself. This can easily be achieved by cross correlating the top and bottom halves of the undistorted image to obtain one vector, then repeating with the left and right halves to obtain another. The image is then tiled and averaged until an area larger than the original image is formed, with an added advantage that this reduces noise in the reference and minimises any small strain that is present. This oversized image is then cross correlated with the undistorted image so that an aligned image can be extracted from the oversized image. This then forms a reference suitable for use within non-linear least squares fitting.

## Strain precision

To verify the strain precision of the flyback corrected image, GPA analysis was performed in pre- and post-corrected images, shown in **Fig S6**. Si was used a known specimen with negligible strain, and before corrected there is a clear compressive strain, as seen in **Fig. S6 (d)**. Post correction this strain has been removed, with average strain values of the left half of the image reducing from $-2.420 \pm 0.004$ % before correction to $-0.309 \pm 0.001$ % afterwards. There is similarly a small decrease in measured strain in the right half of the image, going from $0.136 \pm 0.001$ % to $0.066 \pm 0.001$ %.

## References

HŸTCH, M. J., SNOECK, E. & KILAAS, R. (1998). Quantitative measurement of displacement and strain fields from HREM micrographs. *Ultramicroscopy* **74**, 131–146.

JONES, L., YANG, H., PENNYCOOK, T. J., MARSHALL, M. S. J., VAN AERT, S., BROWNING, N. D., CASTELL, M. R. & NELLIST, P. D. (2015). Smart Align—a new tool for robust non-rigid registration of scanning microscope data. *Advanced Structural and Chemical Imaging* **1**, 8.

PETERS, J. J. P. (2021). *JJPPeters/Strainpp: v1.7*. https://zenodo.org/record/5020766.

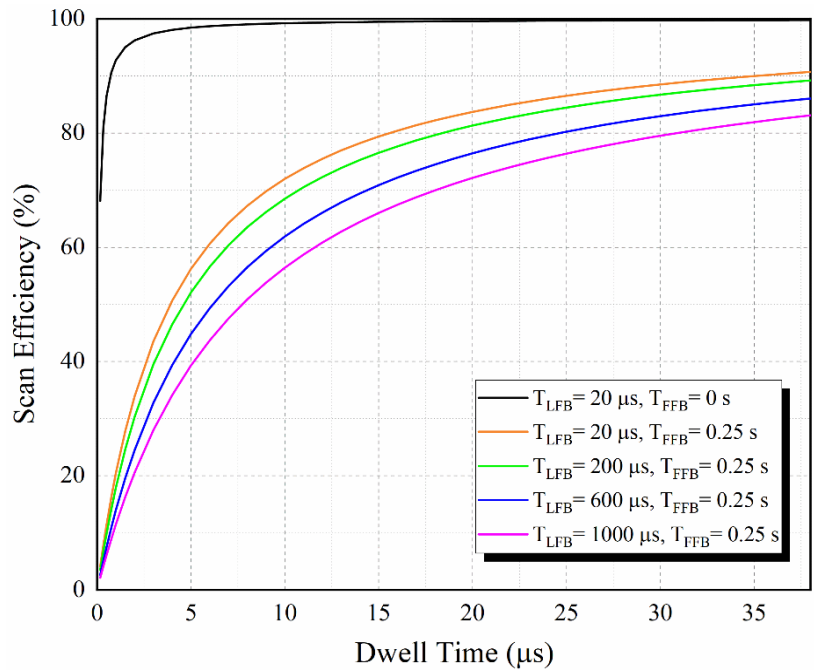

**Figure S1.** *Plots of dose-efficiency as a function of pixel dwell-time for varying flyback settings for a 256 × 256 image. The uppermost line represents the efficiencies which could be achieved with the elimination of frame flyback time. The efficiencies in this case are all lower than for the corresponding settings for a 512 × 512 image.*

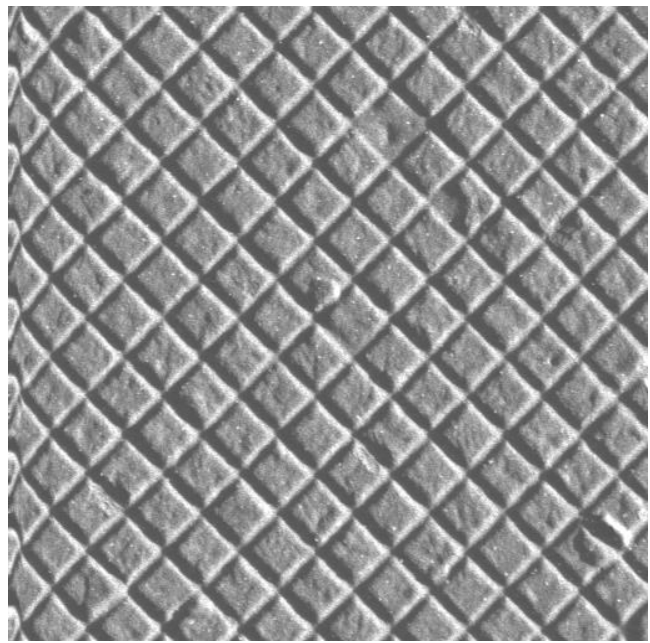

**Figure S2.** *Image captured of a standard cross grating test specimen on a ThermoFisher Titan G2 80-300 kV STEM which displays a compression artefact of the same exponential form as is seen in the images captured on the Nion UltraSTEM 200.*

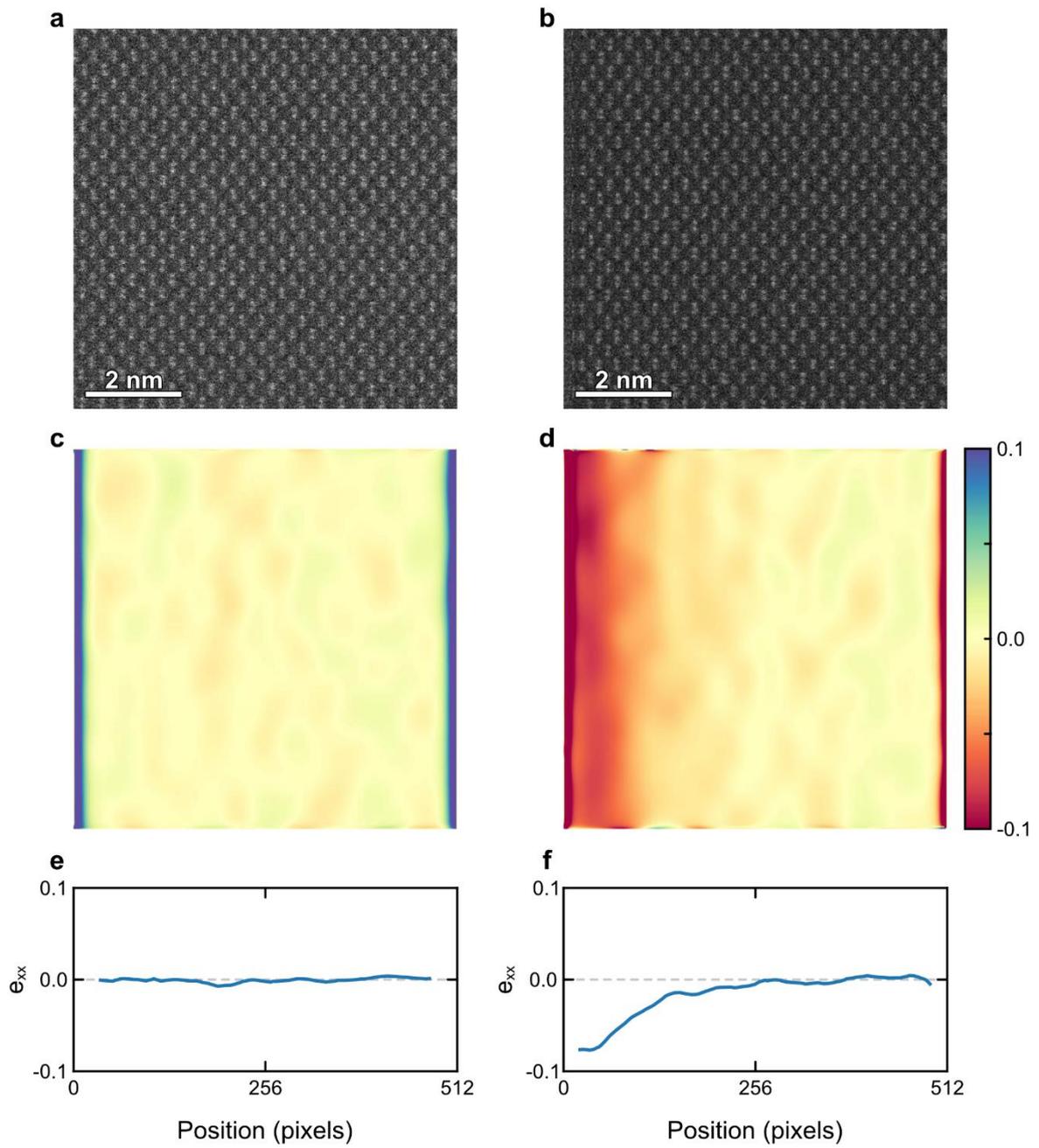

**Figure S3.** *High angle annular dark field image of Si <110> for (**a**) 1 ms flyback time and (**b**) 20 μs flyback time. Both images are the sum of 5 frames, use a dwell time of 2 μs and are 512 × 512 pixels in dimension. (**c**) and (**d**) show $e_{xx}$ distortion maps from (**a**) and (**b**) respectively. (**e**) and (**f**) show profiles of the median distortion from (**c**) and (**d**), respectively, where the edge artefacts have been excluded.*

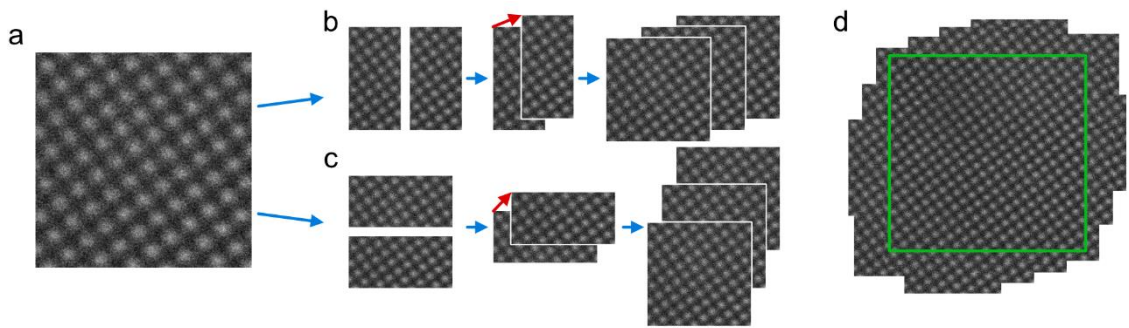

**Figure S4.** *(a) shows a starting undistorted image. The image is split horizontally, (b), and vertically, (c), and a tiling vector (red arrow) is found from cross correlations for self-tiling. (d) shows the resulting oversized image with an area (green box) shown for use as a reference.*

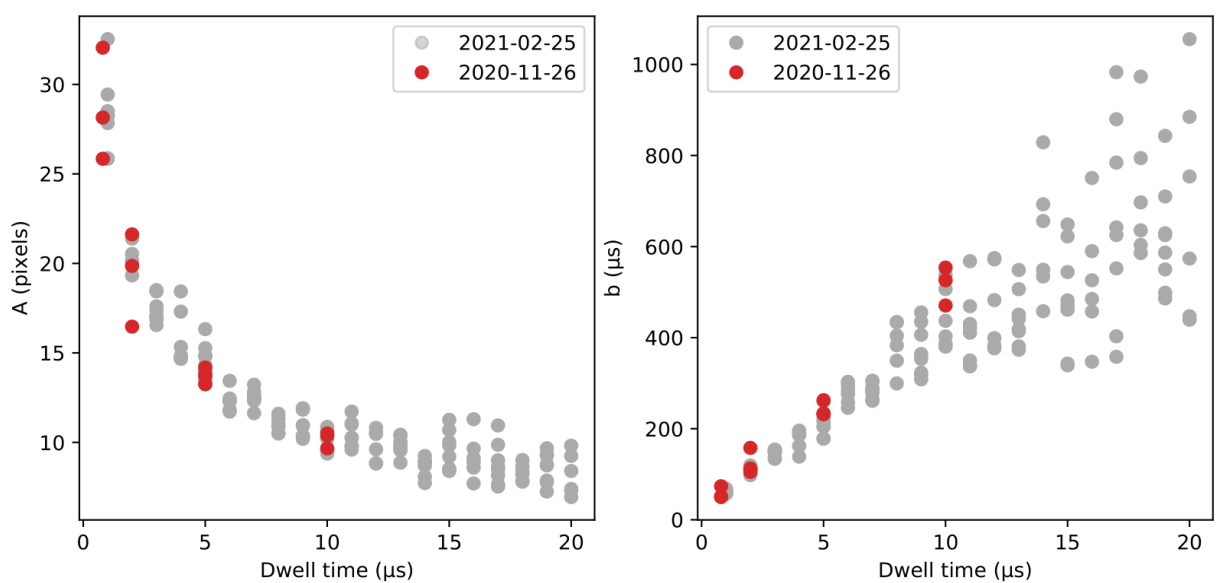

**Figure S5.** *The data from **Figure 5** separated by the date the data was captured. The same trend can be seen three months apart, indicating that the diagnosis performed here remains stable over long periods of time and should not need to be repeated regularly. At longer dwell times the compression artefact is confined to fewer pixels and so obtaining a good fit becomes more difficult, and so we see a larger scatter in the data. This contrasts with short dwell times where the compression can affect up to half of the image. As more crystallography appears deformed the fitting algorithm produces a better fit, and we see less scatter in the data. Thankfully we see better fits with shorter dwell times as this is when the image is most affected. At longer dwell times the compression is confined to very few pixels, and although the variation may seem worrying it has little effect on the correction, and the compression artefact has been reliably corrected under these conditions.*

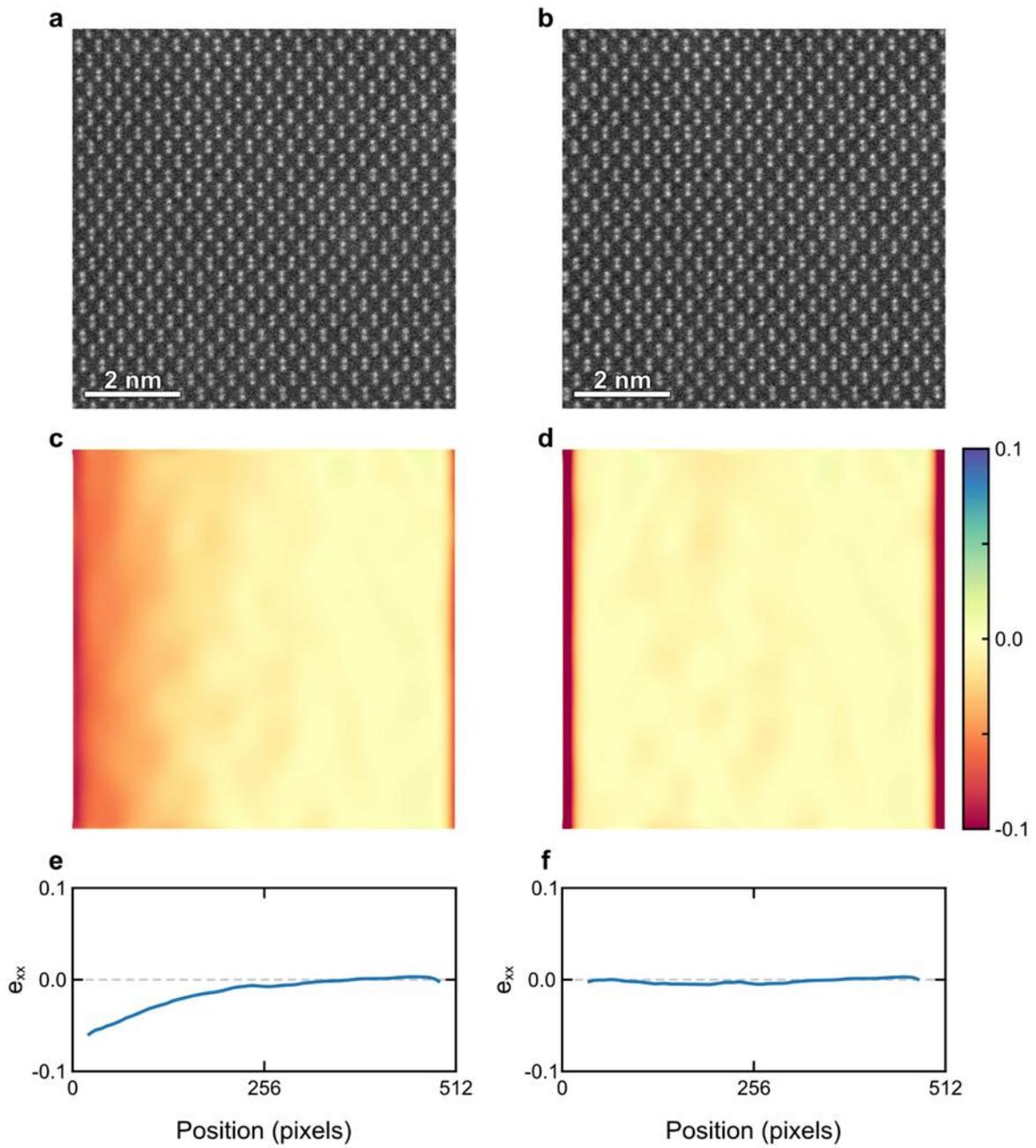

**Figure S6.** *High angle annular dark field images of Si <110> for with a dwell time of 1 μs, a flyback of 60 μs and an image size of 512 × 512 pixels. (**a**) and (**b**) show the image pre and post flyback correction, respectively. (**c**) and (**d**) show $e_{xx}$ distortion maps from (**a**) and (**b**) respectively. (**e**) and (**f**) show profiles of the median distortion from (**c**) and (**d**), respectively, where the edge artefacts have been excluded.*

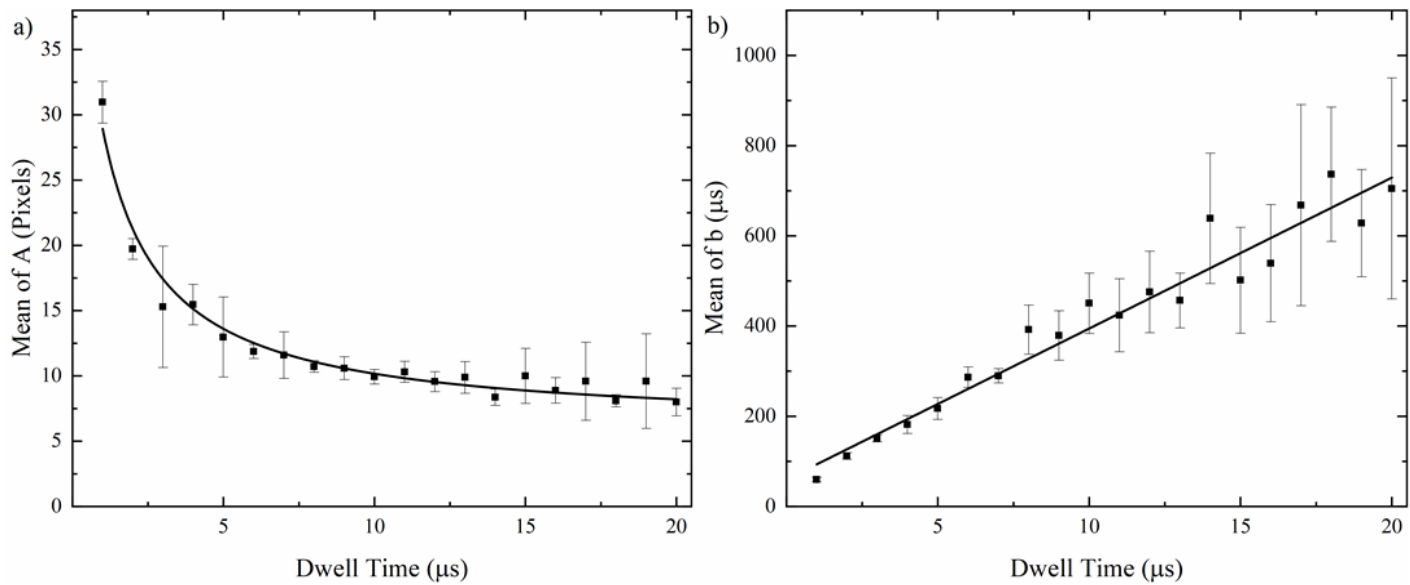

**Figure S7.** *The mean and standard deviation of the data in* **Figure 5** *as a function of dwell time, with the same fit lines overlaid.*

**Table S1.** *Error estimates and goodness-of-fit parameters for the fitted lines shown in* **Figure 5**, *here parametrised by u, v, and w. The observed linear relationship between 'b' and dwell time shows that their ratio can be considered a constant of the system.*

| Figure | Figure 5 a) | Figure 5 b) |
|---|---|---|
| Equation | $A = u + v/(\delta_t + w)$ | $b = u + v * \delta_t$ |
| u | $6.0 \pm 0.3$ | $60 \pm 18$ |
| v | $45 \pm 5$ | $33 \pm 2$ |
| w | $0.97 \pm 0.34$ | |
| Adj. R-Square | 0.99 | 0.76 |